\documentclass[11pt]{article}
\usepackage{amsmath,amsthm,amsfonts,amssymb,comment}

\renewcommand{\=}{\doteq}
\newcommand{\p}{\partial}
\newcommand{\f}{\dot{f}}
\newcommand{\g}{\dot{g}}
\newcommand{\pp}{\dot{p}}
\newcommand{\s}{\dot{s}}
\newcommand{\RR}{\mathbb{R}}
\newcommand{\al}{\alpha}
\newcommand{\la}{\lambda}
\newcommand{\apr}{\approx}
\newcommand{\F}{\dot{F}}

\newtheorem{thm}{Theorem}
\newtheorem{prop}[thm]{Proposition}
\newtheorem{cor}[thm]{Corollary}

\theoremstyle{definition}
 \newtheorem{defn}[thm]{Definition}
\theoremstyle{definition}
 \newtheorem{exam}[thm]{Example}
\theoremstyle{definition}
 \newtheorem{rem}[thm]{Remark}
\theoremstyle{definition}
 \newtheorem{conrem}[thm]{Concluding remark}

\begin{document}

\title{\Large\bf Plane rotations and Hamilton-Dirac mechanics}
\author{\large Eugen Paal and J\"uri Virkepu
\\  \\
Tallinn University of Technology
\\
Ehitajate tee 5, 19086 Tallinn, Estonia
\\ \\
eugen@edu.ttu.ee and jvirkepu@staff.ttu.ee 
}
\date{}
\maketitle
\thispagestyle{empty}

\begin{abstract}
Canonical formalism for $SO(2)$ is developed.
This group can be seen as a toy model of the Hamilton-Dirac mechanics with constraints.
The Lagrangian and Hamiltonian are explicitly constructed and their physical interpretation are given. The Euler-Lagrange and Hamiltonian canonical equations
coincide with the Lie equations. 
It is shown that the constraints satisfy CCR. 
Consistency of the constraints is checked.
\end{abstract}

\section{Introduction and outline of the paper}

The \emph{quantum groups} are conventionally constructed via deformations (e.g \cite{chapre94}).
But  it is also interesting to consider other methods, e.g canonical and path integral quantizations.
Then one has to construct the Lagrangian and Hamiltonian of a group under consideration.
The crucial idea of such an approach is that the Euler-Lagrange and Hamilton canonical equations must be the Lie equations of the Lie (transformation) group. 

In this paper, the canonical formalism for real plane rotations is developed.  
It is shown that the one-parametric real plane rotation group $SO(2)$ can be seen as a toy model of the 
\emph{Hamilton-Dirac mechanics} with constraints \cite{dirac64}.
The Lagrangian and Hamiltonian are explicitly constructed. The Euler-Lagrange and Hamiltonian equations
coincide with the Lie equations. 
Consistency of the constraints is checked. It is also shown that the constraints satisfy the canonical commutation relations (CCR). 

\section{Lagrangian and Lie equations}

Let $SO(2)$ be the rotation group of the real two-plane $\RR^{2}$ .
Rotation of the plane $\RR^{2}$ by an angle $\al\in\RR$ 
is given by the transformation
\[
\begin{cases}
x'=f(x,y,\al)\=x\cos\al-y\sin\al\\
y'=g(x,y,\al)\=x\sin\al+y\cos\al
\end{cases}
\]
We consider the rotation angle $\al$ as a dynamical variable and the functions $f$ and $g$ as \emph{field variables} for $SO(2)$.
Denote
\[
\f\=\p_\al f, \qquad \g\=\p_\al g
\]
The \emph{infinitesimal coefficients} of the transformation are
\[
\begin{cases}
\xi(x,y)\=\f(x,y,0)=-y\\
\eta(x,y)\=\g(x,y,0)=x
\end{cases}
\]
and the \emph{Lie equations} read
\[
\begin{cases}
\f=\xi(f,g)=-g\\
\g=\eta(f,g)=f
\end{cases}
\]
Our first aim is to find such a Lagrangian $L(f,g,\f,\g)$ that the Euler-Lagrange equations
\[
\dfrac{\p L}{\p f}-\dfrac{\p}{\p\al}\dfrac{\p L}{\p\f}=0,
\qquad
\dfrac{\p L}{\p g}-\dfrac{\p}{\p\al}\dfrac{\p L}{\p\g}=0
\]
correspondingly coincide with the Lie equations.

\begin{defn}[Lagrangian]
The Lagrangian $L$ for $SO(2)$ can be defined by
\[
L(f,g,\f,\g)\=\frac{1}{2}(f\g-\f g)-\frac{1}{2}\left(f^{2}+g^{2}\right)
\]
\end{defn}
\begin{thm}
The Euler-Lagrange equations of $SO(2)$ coincide with its Lie equations.
\end{thm}

\begin{proof}
Calculate
\begin{align*}
\dfrac{\p L}{\p f}
&=\dfrac{\p}{\p f}
     \left[\frac{1}{2}(f\g-\f g)-\frac{1}{2}\left(f^{2}+g^{2}\right)\right]=\frac{1}{2}\g-f\\
\dfrac{\p L}{\p\f}
&=\dfrac{\p}{\p\f}
\left[\frac{1}{2}(f\g-\f g)-\frac{1}{2}\left(f^{2}+g^{2}\right)\right]=-\frac{1}{2}g
\qquad\Longrightarrow\quad
\dfrac{\p}{\p\al}\frac{\p L}{\p\f}=-\frac{1}{2}\g
\end{align*}
from which it follows
\[
\dfrac{\p L}{\p f}-\dfrac{\p}{\p\al}\dfrac{\p L}{\p\f}=0
\quad\Longleftrightarrow\quad
\frac{1}{2}\g-f+\frac{1}{2}\g=0
\quad\Longleftrightarrow\quad
\g=f
\]
Analogously calculate
\begin{align*}
\frac{\p L}{\p g}
&=\dfrac{\p}{\p g}
     \left[\frac{1}{2}(f\g-\f g)-\frac{1}{2}\left(f^{2}+g^{2}\right)\right]=-\frac{1}{2}\f-g\\
\dfrac{\p L}{\p\g}
&=\dfrac{\p}{\p\g}
\left[\frac{1}{2}(f\g-\f g)-\frac{1}{2}\left(f^{2}+g^{2}\right)\right]=\frac{1}{2}f
\qquad\Longrightarrow\quad
\dfrac{\p}{\p\al}\frac{\p L}{\p\g}=\frac{1}{2}\f
\end{align*}
from which it follows
\[
\dfrac{\p L}{\p g}-\dfrac{\p}{\p\al}\dfrac{\p L}{\p\g}=0
\quad\Longleftrightarrow\quad
-\frac{1}{2}\f-g-\frac{1}{2}\f=0
\quad\Longleftrightarrow\quad
\f=-g
\qedhere
\]
\end{proof}

\section{Physical interpretation}

The system of Lie equations is equivalent to the following one:
\[
\ddot{f}+f=0=\ddot{g}+g
\]
The Lagrangian of the latter reads
\[
L(f,g,\f,\g)
\=\frac{1}{2}\left(\f^{2}+\g^{2}\right)-\frac{1}{2}\left(f^{2}+g^{2}\right)
\]
The quantity
\[
T\=\frac{1}{2}\left(\f^{2}+\g^{2}\right)
\]
is the \emph{kinetic energy} of a moving point $(f,g)\in\RR^{2}$, meanwhile 
\[
l\=f\g-g\f
\] 
is its \emph{kinetic momentum} with respect to origin $(0,0)\in\RR^{2}$. 
By using the Lie equations  one can easily check that 
\[
\f^{2}+\g^{2}=f\g-g\f
\]
This relation has a simple explanation  in the kinematics of a rigid body \cite{gold}. The kinetic energy of a point
can be represented via its kinetic momentum as follows:
\[
\frac{1}{2}\left(\f^{2}+\g^{2}\right)=T=\frac{l}{2}=\frac{1}{2}(f\g-g\f)
\]
This relation explains the equivalence of the Lagrangians. 
Both Lagrangians give rise to the same extremals.
But  one must remember that this relation holds only on the extremals, i.e for the given Lie equations.

\section{Hamiltonian and Hamilton equations}

Our aim is to develop canonical formalism for the Lie equations.
According to canonical formalism, define the \emph{canonical momenta} as
\begin{align*}
p&\=\dfrac{\p L}{\p\f}=\dfrac{\p}{\p\f}\left[\frac{1}{2}(f\g-\f g)-\frac{1}{2}\left(f^{2}+g^{2}\right)\right]=-\frac{g}{2}\\
s&\=\dfrac{\p L}{\p\g}=\dfrac{\p}{\p\g}\left[\frac{1}{2}(f\g-\f g)-\frac{1}{2}\left(f^{2}+g^{2}\right)\right]=+\frac{f}{2}
\end{align*}
Note that the canonical momenta do not depend on velocities and so we are confronted with a \emph{constrained system}
with two \emph{constraints}
\[
\varphi_1(f,g,p,s)\=p+\frac{g}{2}=0,
\qquad
\varphi_2(f,g,p,s)\=s-\frac{f}{2}=0
\]

\begin{defn}[Hamiltonian]
According to Dirac \cite{dirac64}, the \emph{Hamiltonian} $H$ for $SO(2)$ can be defined by
\begin{align*}
H
&\=\overbrace{p\f+s\g-L}^{H'}+\la_1\varphi_1(f,g,p,s)+\la_2\varphi_2(f,g,p,s)\\
&=p\f+s\g-L+\la_1\left(p+\frac{g}{2}\right)+\la_2\left(s-\frac{f}{2}\right)
\end{align*}
where $\la_1$ and $\la_2$ are the \emph{Lagrange multipliers}.
\end{defn}

\begin{prop}
The Hamiltonian of $SO(2)$ can be presented as
\[
H=\frac{1}{2}\left(f^{2}+g^{2}\right)+\la_1\left(p+\frac{g}{2}\right)+\la_2\left(s-\frac{f}{2}\right)
\]
\end{prop}

\begin{proof}
It is sufficient to calculate
\begin{align*}
H'
&\=p\f+s\g-L\\
&=p\f+s\g-\frac{1}{2}(f\g-\f g)+\frac{1}{2}\left(f^{2}+g^{2}\right)\\
&=\f\left(p+\frac{g}{2}\right)+\g\left(s-\frac{f}{2}\right)+\frac{1}{2}\left(f^{2}+g^{2}\right)\\
&=\frac{1}{2}\left(f^{2}+g^{2}\right)
\qedhere
\end{align*}
\end{proof}

\begin{thm}[Hamiltonian equations]
If the Lagrange multipliers 
\[
\la_1=-g,\qquad \la_2=f
\]
 then the Hamiltonian equations
\[
\f=\dfrac{\p H}{\p p},
\qquad
\g=\dfrac{\p H}{\p s},
\qquad
\pp=-\dfrac{\p H}{\p f},
\qquad
\s=-\dfrac{\p H}{\p g}
\]
coincide with the Lie equations of $SO(2)$.
\end{thm}

\begin{proof}
Really, first calculate
\begin{align*}
\f
&=\dfrac{\p H}{\p p}
=\dfrac{\p}{\p p}\left[\frac{1}{2}\left(f^{2}+g^{2}\right)-g\left(p+\frac{g}{2}\right)+f\left(s-\frac{f}{2}\right)\right]
=-g\\
\g
&=\dfrac{\p H}{\p s}
=\dfrac{\p}{\p s}\left[\frac{1}{2}\left(f^{2}+g^{2}\right)-g\left(p+\frac{g}{2}\right)+f\left(s-\frac{f}{2}\right)\right]
=f
\end{align*}
Similarly calculate
\begin{align*}
\pp
&=-\dfrac{\p H}{\p f}
=-\dfrac{\p}{\p f}\left[\frac{1}{2}\left(f^{2}+g^{2}\right)-g\left(p+\frac{g}{2}\right)+f\left(s-\frac{f}{2}\right)\right]\\
&=-f-s+f=-s\\
\s
&=-\dfrac{\p H}{\p g}
=-\dfrac{\p}{\p g}\left[\frac{1}{2}\left(f^{2}+g^{2}\right)-g\left(p+\frac{g}{2}\right)+f\left(s-\frac{f}{2}\right)\right]\\
&=-g+p+g=p
\end{align*}
Now use  here the constraints $p=-g/2$ and $s=f/2$ to obtain
\[
\begin{cases}
\pp=-s\\
\s=p
\end{cases}
\Longrightarrow\quad
\begin{cases}
-\frac{1}{2}\g=-\frac{1}{2}f\\
+\frac{1}{2}\f=-\frac{1}{2}g
\end{cases}
\Longrightarrow\quad
\begin{cases}
\g=f\\
\f=-g
\end{cases}
\qedhere
\]
\end{proof}

\begin{rem}
One must  remember that on the constraints must be applied {\bf after} the calculations of the partial derivatives of $H$.
 \end{rem}

\begin{cor}
The Hamiltonian of $SO(2)$ can be presented in the form
\[
H=fs-gp
\]
Then the Hamilton equations coincide with the Lie equations of $SO(2)$.
\end{cor}

\begin{rem}
Note that our hamiltonian $H$ is the  \emph{angular momentum} of the point $(f,g)\in\RR^{2}$. 
This is natural, because we consider plane rotations: the angular momentum is the generator of the rotations.
Hamiltonian obtained from  conventional Lagrangian  will be the \emph{total energy}
\begin{align*}
E
&\=\frac{1}{2}\left(p^{2}+s^{2}\right)+\frac{1}{2}\left(\f^{2}+\g^{2}\right)\\
&=\frac{1}{2}(fs-gp)+\frac{1}{2} \left(\f^{2}+\g^{2}\right)
\end{align*}
\end{rem}

\section{Poisson brackets and constraint algebra}

\begin{defn}[observables and Poisson brackets]
Sufficiently  smooth functions of the canonical varibles are called  \emph{observables}.
The \emph{Poisson brackets} of the observables $F$ and $G$ are defined by
\[
\{F,G\}
\=\dfrac{\p F}{\p f}\dfrac{\p G}{\p p}
-\dfrac{\p F}{\p p}\dfrac{\p G}{\p f}
+\dfrac{\p F}{\p g}\dfrac{\p G}{\p s}
-\dfrac{\p F}{\p s}\dfrac{\p G}{\p g}
\]
\end{defn}

\begin{exam}
In particular, one can easily check that
\[
\{f,p\}=1=\{g,s\}
\]
and all other Poisson brackets between canonical variables vanish.
\end{exam}

\begin{exam}
In particular,
\begin{align*}
\{\varphi_1,H'\}
=\left\{p+\frac{g}{2},H'\right\}
=-\dfrac{\p H'}{\p f}+\frac{1}{2}\dfrac{\p H'}{\p s}
=-\frac{1}{2}\dfrac{\p}{\p f}\left(f^{2}+g^{2}\right)
=-f
\end{align*}
and similarly
\begin{align*}
\{\varphi_2,H'\}
=\left\{s-\frac{f}{2},H'\right\}
=-\frac{1}{2}\dfrac{\p H'}{\p p}-\dfrac{\p H'}{\p g}
=-\frac{1}{2}\dfrac{\p}{\p g}\left(f^{2}+g^{2}\right)
=-g
\end{align*}
\end{exam}

\begin{defn}[weak equality]
The observables  $A$ and $B$ are called \emph{weakly equal}, if 
\[
(A-B)\Big|_{\varphi_1=0=\varphi_2}=0
\] 
In this case we write $A\apr B$.

\end{defn}

\begin{thm}
The Lie equations read 
\[
\f\apr\dfrac{\p H}{\p p},
\qquad
\g\apr\dfrac{\p H}{\p s},
\qquad
\pp\apr-\dfrac{\p H}{\p f},
\qquad
\s\apr-\dfrac{\p H}{\p g}
\]
\end{thm}

\begin{thm}
The Lie equations of $SO(2)$ can be presented in the Poisson-Hamilton form
\[
\f\apr\{f,H\},
\qquad
\g\apr\{g,H\},
\qquad
\pp\apr\{p,H\},
\qquad
\s\apr\{s,H\}
\] 
\end{thm}

\begin{proof}
As an example, check the third equation.
We have
\[
\{p,H\}
\=\dfrac{\p p}{\p f}\dfrac{\p H}{\p p}
-\dfrac{\p p}{\p p}\dfrac{\p H}{\p f}
+\dfrac{\p p}{\p g}\dfrac{\p H}{\p s}
-\dfrac{\p p}{\p s}\dfrac{\p H}{\p g}
=-\dfrac{\p H}{\p f}\apr\pp
\qedhere
\]
\end{proof}

\begin{thm}
The equation of motion of an observable $F$ reads
\[
\F\apr\{F,H\}
\]
\end{thm}

\begin{proof}
By using the Hamilton equations,  calculate
\begin{align*}
\F
&=\dfrac{\p F}{\p f}\f
+\dfrac{\p F}{\p p}\pp
+\dfrac{\p F}{\p g}\g
+\dfrac{\p F}{\p s}\s  \\
&\apr\dfrac{\p F}{\p f}\dfrac{\p H}{\p p}
-\dfrac{\p F}{\p p}\dfrac{\p H}{\p f}
+\dfrac{\p F}{\p g}\dfrac{\p H}{\p s}
-\dfrac{\p F}{\p s}\dfrac{\p H}{\p g}\\
&\=\{F,H\}
\qedhere
\end{align*}
\end{proof}

\begin{thm}[constraint algebra]
The constraints of $SO(2)$ satsify the CCR relations
\[
\{\varphi_1,\varphi_1\}=0=\{\varphi_2,\varphi_2\},
\qquad 
\{\varphi_1,\varphi_2\}=1
\]
\end{thm}

\begin{proof}
First two relations are evident. To check the third one, calculate
\begin{align*}
4\{\varphi_1,\varphi_2\}
&=\{2p+g,2s-f\}\\
&\=\dfrac{\p(2p+g)}{\p f}\dfrac{\p(2s-f)}{\p p}
-\dfrac{\p(2p+g)}{\p p}\dfrac{\p(2s-f)}{\p f}\\
&+\dfrac{\p(2p+g)}{\p g}\dfrac{\p(2s-f)}{\p s}
-\dfrac{\p(2p+g)}{\p s}\dfrac{\p(2s-f)}{\p g}\\
&=-2\dfrac{\p(2s-f)}{\p f}+\dfrac{\p(2s-f)}{\p s}\\
&=2+2=4
\qedhere
\end{align*}
\end{proof}

\section{Consistency}

Now consider the dynamical behaviour of the constraints. 
Note that
\[
\varphi_1=0=\varphi_2
\quad\Longrightarrow\quad
\dot{\varphi_1}=0=\dot{\varphi_2}
\]
To be consistent with equations of motion we  must prove the
\begin{thm}[consistency]
The constraints of $SO(2)$ satisfy equations
\[
\{\varphi_1,H\}\apr\dot{\varphi}_1=0,
\qquad 
\{\varphi_2,H\}\apr\dot{\varphi}_2=0
\]
\end{thm}

\begin{proof}
Really, first calculate
\begin{align*}
\{\varphi_1,H\}
&\=\{\varphi_1,H'+\la_1\varphi_1+\la_2\varphi_2\}\\
&\apr\{\varphi_1,H'\}+\la_1\{\varphi_1,\varphi_1\}+\la_2\{\varphi_1,\varphi_2\}\\
&=-f+\la_1\cdot0+\la_2\cdot1\\
&=-f+f\\
&=0\\
&=\dot{\varphi}_1
\end{align*}
Similarly
\begin{align*}
\{\varphi_2,H\}
&\=\{\varphi_2,H'+\la_1\varphi_1+\la_2\varphi_2\}\\
&\apr\{\varphi_2,H'\}+\la_1\{\varphi_2,\varphi_1\}+\la_2\{\varphi_2,\varphi_2\}\\
&=-g-\la_1\cdot1+\la_2\cdot0\\
&=-g+g\\
&=0\\
&=\dot{\varphi}_2
\qedhere
\end{align*}
\end{proof}

\begin{conrem}
Once the canonical structure of $SO(2)$ established, one can perform the canonical quantization
as well.
This actually means the quantization of the angular momentum.
\end{conrem}

\section*{Acknowledgement}

The paper was in part supported by the Estonian SF Grant 5634.


\begin{thebibliography}{9}
\itemsep-2pt

\bibitem{chapre94}
V.~Chari and A.~Pressley.
A Guide to Quantum Groups.
Cambridge Univ Press, 1994.

\bibitem{gold}
H. Goldstein.
Classical mechanics.
Addison-Wesley Press, Cambridge, 1953.

\bibitem{dirac64}
P.~Dirac. 
Lectures on Quantum Mechanics. 
Yeshiva Univ, New York, 1964.
\end{thebibliography}
\end{document}